\RequirePackage{lineno} 

\documentclass[twocolumn,showpacs,preprintnumbers,amsmath,amssymb,nofootinbib]{revtex4}

\usepackage{graphicx}
\usepackage{dcolumn}
\usepackage{bm}
 
\usepackage{epstopdf}
\def\bea{\begin{eqnarray}}
\def\eea{\end{eqnarray}}

\def\pp{\mbox{$p$-$p$}}

\def\pa{\mbox{$p$-A}}

\def\auau{\mbox{Au-Au}}

\def\pbpb{\mbox{Pb-Pb}}
\def\ppb{\mbox{$p$-Pb}}
\def\pn{\mbox{$p$-N}}
\def\aa{\mbox{A-A}}
\def\nn{\mbox{N-N}}

\def\pt{$p_t$}

\def\yt{$y_t$}

\def\mmpt{$\bar p_t$}

\begin{document} 

\setlength{\pdfpagewidth}{8.5in}
\setlength{\pdfpageheight}{11in}

\setpagewiselinenumbers
\modulolinenumbers[5]

\preprint{version 0.2}

\title{Exclusivity of p-N interactions within p-A collisions
}

\author{Thomas A.\ Trainor}\affiliation{CENPA 354290, University of Washington, Seattle, WA 98195}


\date{\today}

\begin{abstract}
A geometric Glauber model applied to \pp\ or \pa\ collisions assumes that participant partons within a projectile nucleon independently follow eikonal trajectories, just as for projectile nucleons within target nuclei,  and interact independently with any target nucleons encountered along their trajectory. A noneikonal quadratic relation for dijet production observed for \pp\ collisions suggests that such assumptions for \pn\ collisions may be incorrect. Data from \pa\ collisions supports that conclusion. For both isolated \pp\ collisions and for \pn\ interactions within \pa\ collisions a proton projectile interacts with only a {\em single} target nucleon within a brief time interval. Multiple {\em overlapping} \pn\ collisions are inhibited, suggesting the {\em exclusivity} of \pn\ interactions within \pa\ collisions. 
\end{abstract}


\maketitle

 \section{Introduction}

A geometric Glauber Monte Carlo (MC) model applied to composite A-B nuclear collisions assumes that constituents $x$  of A follow straight-line trajectories through partner B according to the eikonal approximation and interact (collide) with some fraction of constituents $y$ in B (and the reverse) according to some cross section $\sigma_{xy}$. Constituents that have interacted at least once are {\em participants}. Participants must lie within an A-B overlap region depending on A-B impact parameter $b$, and each participant in A may interact with partners in B only within its own {\em eikonal corridor} defined by $\sigma_{xy}$. The geometric Glauber MC has been applied to nucleon constituents within nucleus-nucleus (\aa) collisions~\cite{glauber} and to parton (quantum chromodynamic quark or gluon) constituents within nucleon-nucleon (\nn) collisions~\cite{pythia}.

Recently, a geometric Glauber MC was applied to 5 TeV \ppb\ collisions~\cite{aliceppbprod} to predict \ppb\ centrality (e.g.\ impact parameter $b$ or nucleon participant number $N_{part}$) vs some observable quantity (e.g.\ hadron mean charge density $\bar \rho_0 = n_{ch} / \Delta \eta$ within some pseudorapidity acceptance $\Delta \eta$). The Glauber-inferred relation $N_{part}(\bar \rho_0)$ is very different from one derived from a study in Ref.~\cite{tommpt} of ensemble-mean transverse momentum \mmpt\ data from the same collision system~\cite{alicempt} based on a two-component (soft + hard) model (TCM) of hadron production near midrapidity ($\eta \approx 0$) in high-energy nuclear collisions. The TCM-based study reviewed analysis methods and resulting differences and suggested that the Glauber model as typically applied to A-B collisions is not a valid description of asymmetric \pa\ collisions. In particular, conventional assumptions leading to assignment of certain nucleons as participants based on the possibility of multiple {\em simultaneous} \pn\ interactions of a single projectile proton may be questioned. The alternative is {\em exclusivity}, in which a projectile proton can interact (collide) with only a single target nucleon within some time interval.

For purposes of this letter specific vocabulary is required to distinguish between \pn\ {\em encounters} determined by a geometric model and effective (e.g.\ inelastic) \pn\ {\em collisions} that result in significant involvement of constituent partons. According to the exclusivity hypothesis participants arise only from {\em isolated} inelastic collisions.

\section{The TCM and $\bf p$-$\bf p$ collisions}

In the TCM context parton constituents in projectile nucleons may produce hadrons within composite A-B collisions by longitudinal fragmentation to neutral hadron pairs (soft) or by transverse fragmentation (of large-angle-scattered partons) to jets (hard).  A specific hadron production model must specify which partons in which nucleons may interact with which partners. Soft and hard components, as distinguished by specific manifestations in spectra~\cite{ppprd,hardspec,jetspec2,ppquad} and correlations~\cite{porter2,porter3,anomalous,ppquad}, are represented respectively by charge densities $\bar \rho_s$ and $\bar \rho_h$, with total hadron charge density $\bar \rho_0 = \bar \rho_s + \bar \rho_h$.


Information about \nn\  collision mechanisms may be derived from isolated  proton-proton (\pp) collisions and from \pa\ (e.g.\ \ppb) data. The two sources provide complementary evidence for \pn\ exclusivity. Hadron production in high-energy \pp\ collisions is observed to fluctuate dramatically~~\cite{aliceppmult}. The source of fluctuations may be one of two limiting cases: 

(a) A conventional model of composite A-B collisions based on the eikonal approximation~\cite{glauber} is applied to \pp\ collisions~\cite{pythia}. Impact parameter $b$ between colliding protons,  each with internal structure represented by a parton distribution function (PDF), is random.  Only partons within fluctuating \pp\ overlap regions (participants) interact and, for each parton, only with partners within an {eikonal corridor} along its trajectory determined by a partonic cross section. Fluctuations in $b$ and  proton structure produce fluctuations in hadron production.

(b) Alternatively, what fluctuates in \pp\ collisions is the depth of splitting cascades on momentum fraction $x$ that produce participant partons within each proton. The proton PDF is only a mean-value representation of fluctuating cascades. In a given collision each participant parton from one proton can interact with {\em any} participant in the partner proton. There is no geometric exclusion because impact parameter is not relevant and the eikonal approximation is apparently not valid for \pp\ collisions.

If the eikonal approximation were relevant to \pp\ collisions $\bar \rho_h \propto \bar \rho_s^{4/3}$ should apply, by analogy with the equivalent for nucleus-nucleus (\aa) collisions---the number of binary \nn\ collisions $N_{bin} \approx (N_{part}/2)^{4/3}$~\cite{powerlaw}---with \pp\ participant partons represented by $\bar \rho_s \sim N_{part}$ and dijets represented by $\bar \rho_h \sim N_{bin}$.  Instead, quadratic relation $\bar \rho_h \approx \alpha \bar \rho_s^2$ represents \pp\ spectrum and correlation data accurately over a large range of midrapidity charge densities~\cite{ppprd,ppquad}. Relative to a non-single-diffractive (NSD) \pp\ mean value $\bar \rho_{0NSD}$, charge densities $\bar \rho_0$ varying from a fraction of $\bar \rho_{0NSD}$ up to $\approx 10\,  \bar \rho_{0NSD}$  correspond to approximately {\em 100-fold increase} in dijet production~\cite{ppprd}. 
The noneikonal quadratic relation manifested by \pp\ data suggests that {\em any} participant parton, as represented by $\bar \rho_s$, may interact within a \pp\ collision to produce MB dijets, as represented by $\bar \rho_h$. Collision partners are not restricted to an eikonal corridor nor are parton participants confined to a limited \pp\ overlap region. Scenario (b) is thus strongly supported by \pp\ data and (a) is rejected, suggesting that \pp\ collisions follow an {\em all or nothing} strategy for constituent interactions.

\section{Evidence from $\bf p$-$\bf Pb$ data}

Further information on \nn\ collision properties is derived from recent analysis of  ensemble mean \mmpt\ and \pt\ spectrum data from \ppb\ collisions. TCM analysis of \mmpt\ data from \ppb\ collisions permits inference of the relation between $N_{part}$ and $\bar \rho_0$ for average \pn\ collisions within \ppb\ collisions. Centrality analysis of \ppb\ collisions based on a geometric Glauber model provides competing predictions for the same relationship, and large differences emerge.

 The TCM for A-B $\bar p_t'$ with a low-\pt\ acceptance cut is
\bea \label{pampttcm1}
  \bar p_t' ~&\approx&~ \frac{\bar p_{ts} + x(n_s) \nu(n_s) \, \bar p_{thNN}(n_s)}{\xi + x(n_s)\, \nu(n_s)},
\eea
where $\xi \approx 0.75$ is the soft-component fraction admitted by the acceptance cut, $x(n_s) \equiv \bar \rho_{hNN} / \bar \rho_{sNN}$ and $\nu \equiv 2 N_{bin} / N_{part}$.
For \pa\ data evolution of factors $x(n_s)\, \nu(n_s)$ from strictly \pp--like to alternative behavior is observed near a transition point  $\bar \rho_{s0}$, but $\bar p_{thNN}(n_s) \rightarrow \bar p_{th0}$ is assumed fixed for \pa\ (no jet modification)~\cite{tommpt}. Factor $x(n_s)$ is modeled as a simple extrapolation of \pp\
\bea \label{xmodel}
x(n_s) &=& \frac{\alpha}{\left\{[1/ \bar \rho_s]^{n_1} + [1/f(n_s)]^{n_1}\right\}^{1/n_1}},
\eea
where $f(n_s) = \bar \rho_{s0} + m_0(\bar \rho_s - \bar \rho_{s0})$ and $n_1 \approx 5$.
Below the transition near $\bar \rho_{s0} \approx 15$, $x(n_s) \approx \alpha \bar \rho_s$ as for \pp\ collisions~\cite{ppquad}. Above the transition $x(n_s) \approx m_0 \alpha \bar \rho_s$ with $m_0 \approx 0.1$. Given $x(n_s)$ $N_{part}/2 = \alpha \bar \rho_s / x(n_s) =1/[2 - \nu(n_s)]$ completes the TCM \ppb\ centrality model~\cite{tommpt}.

\begin{figure}[h]
  \includegraphics[width=1.66in]{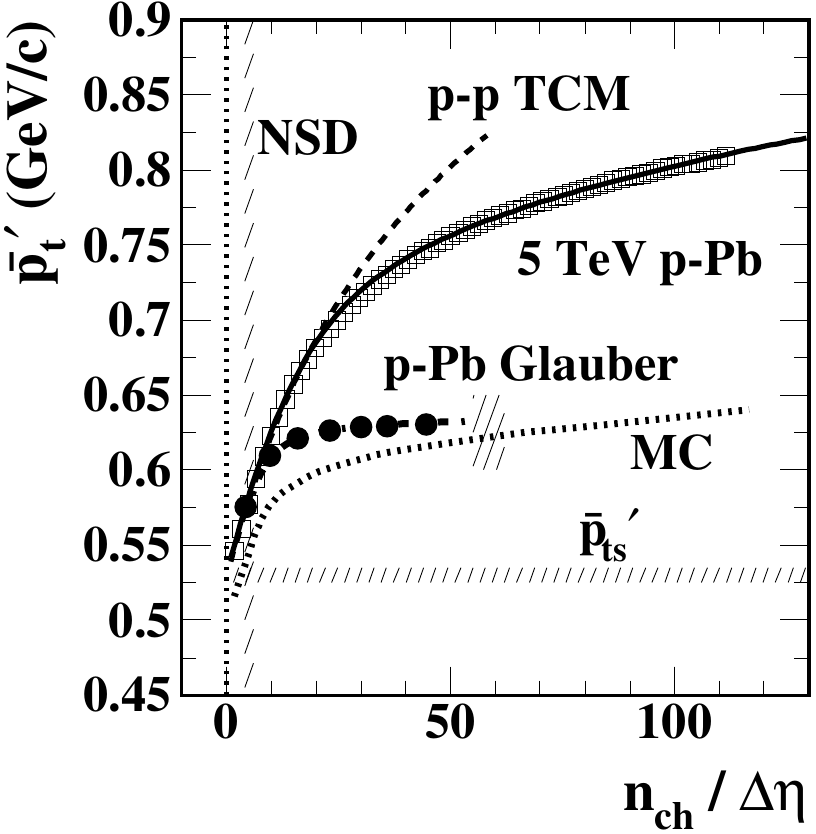}
 \includegraphics[width=1.65in]{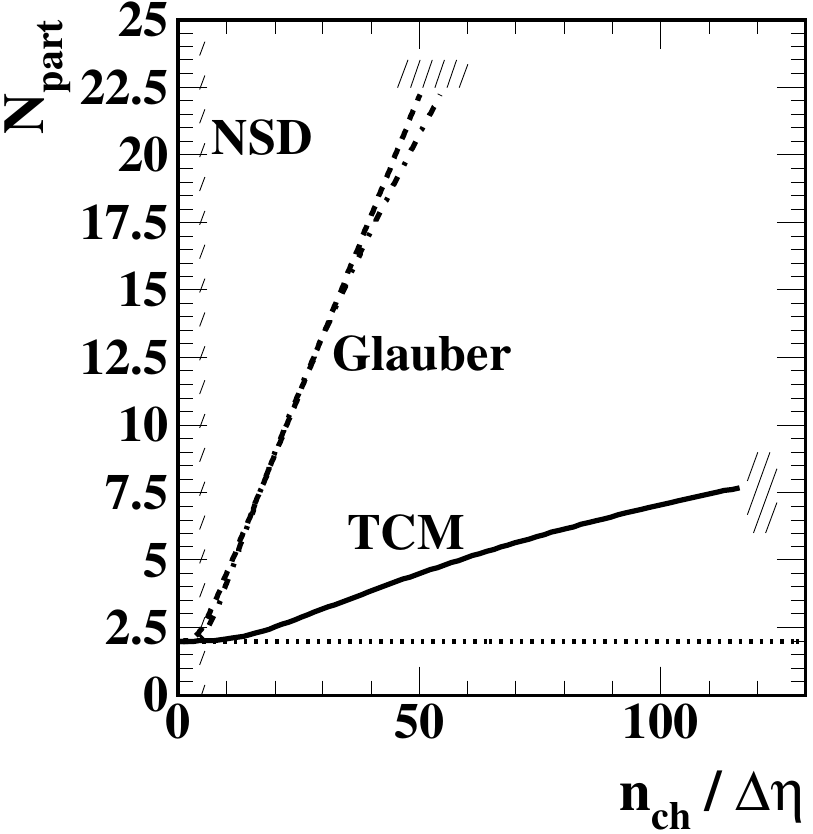}
  \caption{\label{ppbmultx}
  Left:   Uncorrected ensemble-mean \pt\ or $\bar p_t'$ vs corrected $\bar \rho_0 = n_{ch} / \Delta \eta$ for 5 TeV \ppb\ collisions from Ref.~\cite{alicempt} (open squares). The solid curve is the \ppb\ TCM. The solid points are predictions derived from the Glauber centrality analysis in Ref.~\cite{aliceppbprod}.  The Glauber MC curve (dotted) is taken from Fig.~3 of Ref.~\cite{alicempt}.
  Right:  $N_{part}$ vs $\bar \rho_0$ for the Glauber analysis of Ref.~\cite{aliceppbprod} (dash-dotted), for ideal $N_{part}$ scaling (dashed) and for the \ppb\ \mmpt\ TCM (solid). 
   } 
 \end{figure}

Figure~\ref{ppbmultx} (left) shows uncorrected $\bar p_t'$ data vs corrected charge density $\bar \rho_0$~\cite{alicempt}. The solid curve is Eq.~(\ref{pampttcm1}) and the limiting case for \pp\ collisions is the dashed curve. The two curves and data coincide precisely up to a transition point near $\bar \rho_0 \approx 20$~\cite{tommpt}. $\bar p_{ts}' = \bar p_{ts}/ \xi = 0.4 / 0.75$ GeV/c is the uncorrected $\bar p_t'$ TCM soft component. The solid dots are predicted values of $\bar p_t'$ for seven \ppb\ centrality classes from 80-100\% to 0-5\% based on the geometric Glauber analysis of Ref.~\cite{aliceppbprod}. The dotted curve labeled MC represents HIJING~\cite{hijing} and AMPT~\cite{ampt} predictions for \ppb\ collisions from Ref.~\cite{alicempt}. The large difference between \ppb\ TCM and \mmpt\ data vs Glauber-based MC predictions for the same system is the main result.

Figure~\ref{ppbmultx} (right) shows \ppb\ centrality ($N_{part}$ vs $\bar \rho_0$) estimates for 5 TeV \ppb\ from the TCM (solid)~\cite{tommpt} and geometric Glauber MC (dash-dotted)~\cite{aliceppbprod}. The dashed curve represents the assumption that $n_{ch} \propto N_{part}$ for \pa\ collisions~\cite{aliceppbprod}. Hatched bands represent limiting values based on statistical significance. Whereas the Glauber MC sets an upper limit $\bar \rho_0 \approx 55$ corresponding to $N_{part} \approx 22$, the TCM description is consistent with \ppb\ \mmpt\ data from Ref.~\cite{alicempt} extending up to $\bar \rho_0 \approx 115$ with $N_{part} \approx 8$.

\begin{figure}[h]
  \includegraphics[width=3.3in]{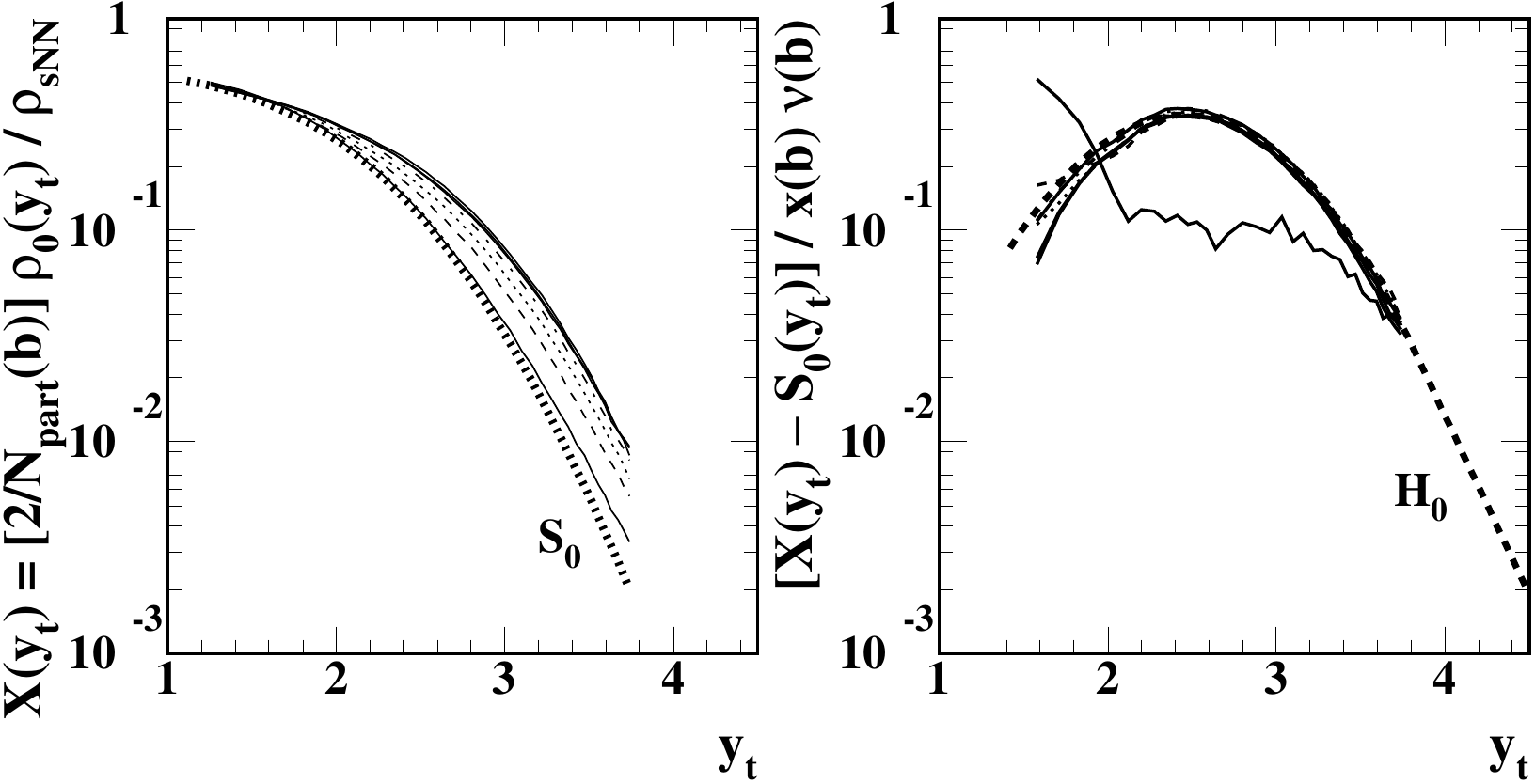}
  \caption{\label{qppb1}
  Left: Identified-pion spectra for 5 TeV \ppb\ collisions from Ref.~\cite{alicepionspec} transformed to \yt\ with Jacobian $m_t p_t / y_t$ and normalized by TCM quantities (thinner curves of several styles). $\hat S_0(y_t)$ is the TCM soft-component model.
  Right: Difference $X(y_t) - \hat S_0(y_t)$ normalized by $x(b) \nu(b) = \alpha \bar \rho_{sNN} \nu(b)$ using TCM values reported in Ref.~\cite{tommpt} (thinner curves). The dashed curve is TCM hard-component model $\hat H_0(y_t)$.
   }  
 \end{figure}

The \ppb\ TCM can be further tested by comparisons with spectrum data. The TCM for A-B spectra is~\cite{tommpt}

\bea \label{ppspectcm1}
\bar \rho_0(y_t,\bar \rho_0) &=&\frac{N_{part}}{2} \bar \rho_{sNN} \hat S_0(y_t) + N_{bin} \bar \rho_{hNN} \hat H_0(y_t),~~
\eea
where $ \hat S_0(y_t)$ and $\hat H_0(y_t) $ are fixed model functions.
Figure~\ref{qppb1} (left) shows identified-pion \pt\ spectra from the same \ppb\ collision system~\cite{alicepionspec} transformed to transverse rapidity \yt\ as in Ref.~\cite{ppquad}.  The spectra are normalized by soft-component density $\bar \rho_s = (N_{part}/2) \bar \rho_{sNN}$ with parameter values from Ref.~\cite{tomglauber}. According to Eq.~(\ref{ppspectcm1}) normalized spectra $X(y_t)$ can be compared with TCM soft-component model $\hat S_0(y_t)$ as defined in Ref.~\cite{alicetomspec}.

Figure~\ref{qppb1} (right) shows difference $X(y_t) - \hat S_0(y_t)$ normalized by $x(n_s) \nu(n_s) = \alpha \bar \rho_{sNN} \nu(n_s)$ using values from Ref.~\cite{tommpt}, with no changes to accommodate these data. According to Eq.~(\ref{ppspectcm1}) the result should be comparable with the TCM hard-component model $\hat H_0(y_t)$, with parameters  $(\bar y_t,\sigma_{y_t},q) = (2.65,0.59,3.9)$ for 5 TeV \pp\ collisions as reported in  Ref.~\cite{alicetomspec}. The dashed curve is $\hat H_0(y_t)$ with $(\bar y_t,\sigma_{y_t},q) \rightarrow (2.45,0.605,3.9)$. The shift of model centroid $\bar y_t$ to a lower value for pions is expected based on fragmentation functions for identified hadrons as in Ref.~\cite{eeprd}.
The TCM describes \ppb\ spectra accurately. 


The \ppb\ centrality analysis in Ref.~\cite{aliceppbprod} is based on a geometric Glauber MC simulation of the differential cross-section distribution on $N_{part}$ which can be compared with the inferred distribution from the TCM.

\begin{figure}[h]
  \includegraphics[width=1.65in]{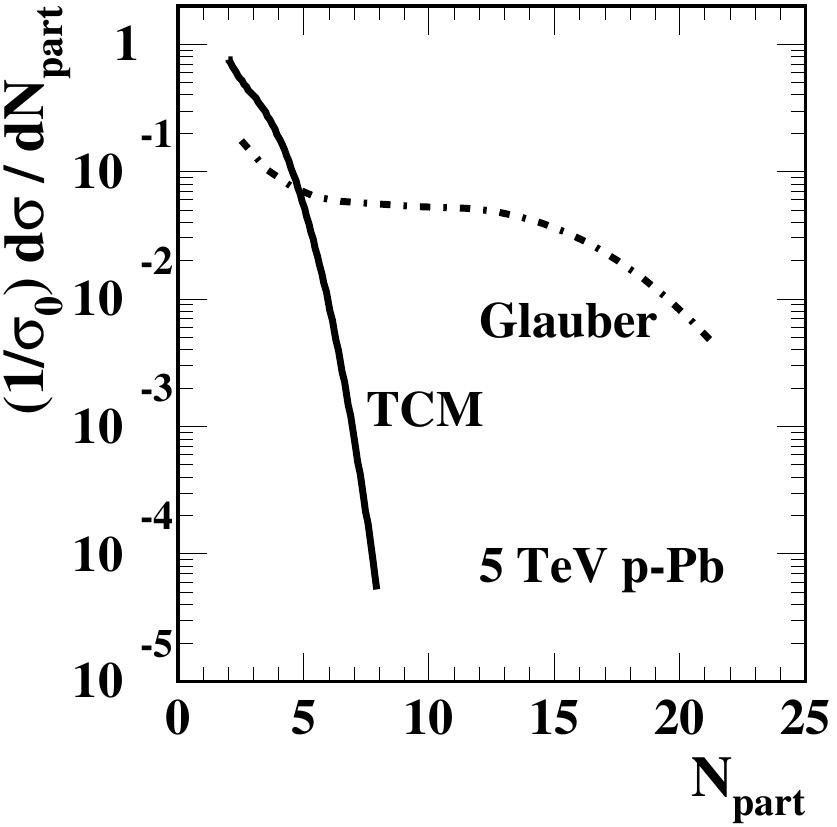}
 \includegraphics[width=1.65in]{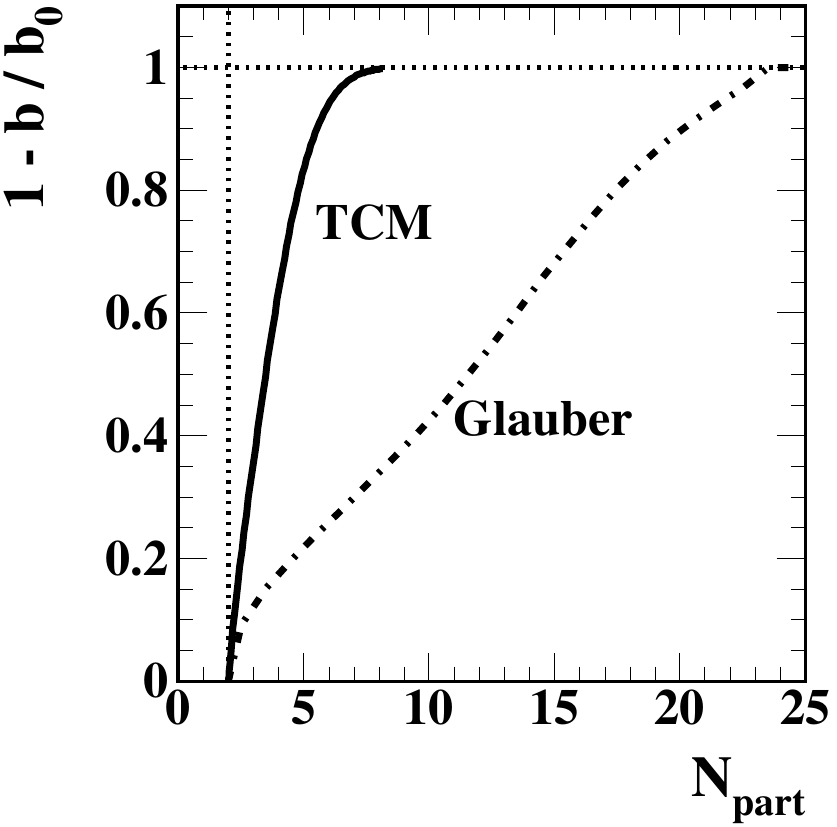}
  \caption{\label{glauber22}
  Left: Differential cross section on $N_{part}$ determined by a geometric Glauber MC (dash-dotted)~\cite{aliceppbprod} and by a TCM analysis of \mmpt\ data (solid)~\cite{tomglauber}.
  Right: Running integrals $1 - \sigma / \sigma_0$ of curves on the left converted to $1 - (b/b_0 = \sqrt{ \sigma / \sigma_0})$.
   } 
 \end{figure}

Figure~\ref{glauber22} (left) shows the geometric-Glauber differential cross section on $N_{part}$ from Ref.~\cite{aliceppbprod} (dash-dotted) extending beyond $N_{part} = 20$. The solid curve is the corresponding TCM cross section inferred from \mmpt\ data as described in Ref.~\cite{tomglauber}. The TCM result suggests that the true maximum $N_{part}$ for central \ppb\ collisions is near 8. Those centrality trends are consistent with Fig.~\ref{ppbmultx} (right).


Figure~\ref{glauber22} (right) shows impact-parameter $b / b_0 = \sqrt{\sigma / \sigma_0}$ trends on $N_{part}$, where $\sigma / \sigma_0$ is the running integral of a differential cross section in the left panel. Whereas the TCM centrality trend attains  its $b/b_0 \rightarrow 0$ limit  near $N_{part} = 8$ the Glauber MC cross section is not fully integrated on $N_{part}$ ($b/b_0 \rightarrow 0$) until $N_{part} \approx 24$. The disagreement between $N_{part}(b)$ trends is the central issue for this letter. Further information can be obtained by examining the \ppb\ geometric Glauber MC in detail.

\section{Geometric Glauber Monte Carlo}


A \ppb\ Glauber MC can be based on the following data: The 5 TeV \ppb\ cross section is $\sigma_0 = \pi b_0^2 \approx 2$ barns = 200 fm$^2$ or $b_0 \approx 8$ fm $\approx (7.1 + 0.85)$ fm (Pb + proton radii). A Pb diameter is spanned by about 8 tangent nucleons. The 5 TeV \pp\ inelastic  cross section is $\sigma_{pp} = \pi b_{0pp}^2 \approx 70$ mb = 7 fm$^2$ or $b_{0pp} \approx1.5$ fm. The Pb nuclear volume is $V_\text{Pb} \approx 1500$ fm$^3$ and the mean nucleon density is $208 / 1500 \approx 0.14$/fm$^3$.  To simplify the MC nucleons are randomly distributed within a cube of side 16 fm maintaining mean density 0.14/fm$^3$. Nucleons outside a Pb nucleus (sphere of radius 7.1 fm) are discarded.  Simulated nuclei then contain 208 nucleons on average. 

A flux of protons is simulated by trajectories randomly distributed across the x-y plane. For a given projectile impact point its eikonal corridor is defined by a circle of radius 1.5 fm consistent with $\sigma_{pp}  \approx 70$ mb. Any nucleon center within that corridor marks an ``encounter.'' It's {\em participant} status depends on the model: For the geometric Glauber MC all encounters are participants.

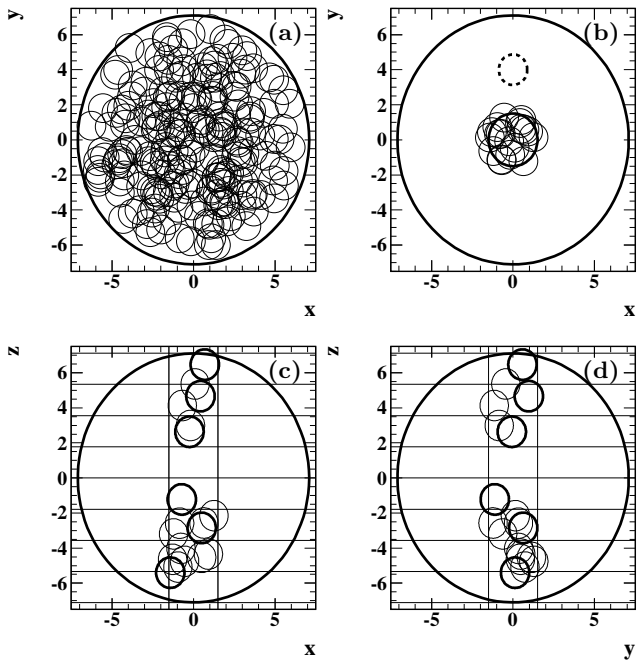
\begin{figure}[h]
  \includegraphics[width=3.3in]{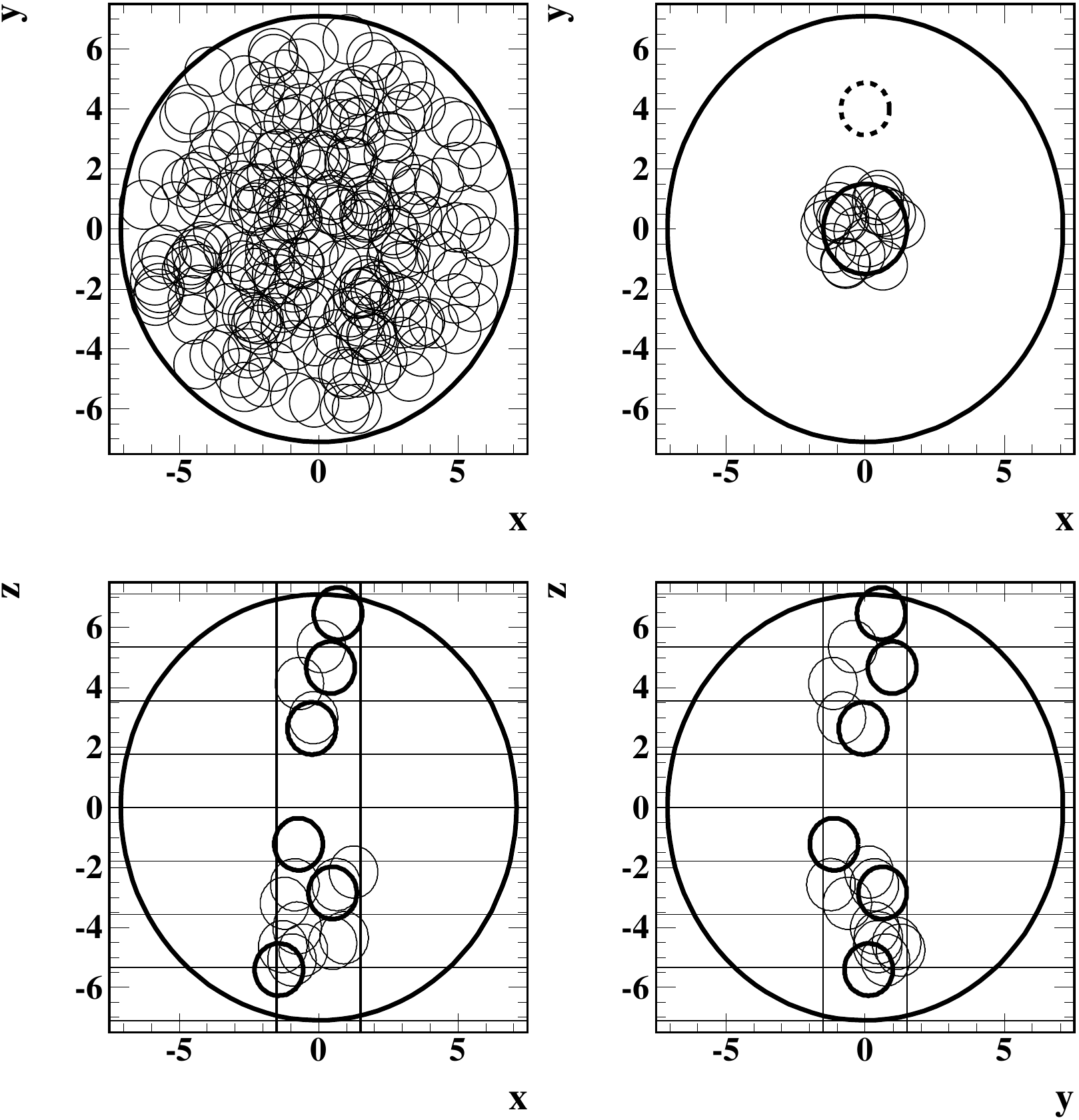}
\put(-140,235) {\bf (a)}
\put(-21,235) {\bf (b)}
\put(-140,107) {\bf (c)}
\put(-21,107) {\bf (d)}
  \caption{\label{cube}
(a): MC-simulated Pb nucleus with mean density  0.14/fm$^3$ and radius 7.1 fm (bold circle).
(b): View along $z$ for a projectile proton with zero impact parameter. The bold circle denotes its eikonal corridor with radius 1.5 fm. The light circles are nucleons with centers within the corridor. The dashed circle is representative, with radius 0.85 fm.
(c) and (d): Side views of the eikonal corridor with included nucleons (encounters). Bold circles are true participants (collisions) corresponding to exclusivity condition $\Delta z_\text{off} = 1.4$ fm.
  } 
\end{figure}

Figure~\ref{cube} (a) shows a simulated Pb nucleus with randomly distributed nucleons. The large-circle radius is 7.1 fm and the nucleon radii are 0.85 fm. A random distribution leads to large density fluctuations including significant voids. Figure~\ref{cube} (b) shows the view along $z$ for a projectile with $b \approx 0$. The bold circle with radius 1.5 fm denotes the eikonal corridor. The dashed circle with radius 0.85 fm represents a target nucleon. The thin circles are nucleons within the eikonal corridor (encounters).

Figs.~\ref{cube} (c) and (d) are side views of the eikonal corridor showing {encountered} target nucleons (circles). The horizontal lines mark eight slices about 1.8 fm thick. The purpose of this simulation is to determine the effect of \pn\ exclusivity---the hypothesis that a projectile proton may interact with only one target nucleon ``at a time,'' where that expression should reflect observations. In this simulation once a projectile proton has collided with one nucleon (upon entering the Pb volume) it cannot collide with another nucleon before moving along $z$ some minimum distance $\Delta z_\text{off}$. For example, in Fig.~\ref{cube} (c) and (d) the actual participants (collisions rather than encounters) for $\Delta z_\text{off} = 1.4$ fm are denoted by bold circles. For this simulation there are 19 encounters (total circles) vs 6 collisions (bold circles) and six occupied slices for a central \ppb\ collision. Note that {\em ten} overlapping non-participant encounters occur within the first two occupied slices.




For each simulated Pb nucleus about 72 projectile protons are  randomly distributed on x-y within the Pb perimeter. Each projectile is propagated through the encountered nucleons within its eikonal corridor. The sum of those encounters (plus the projectile) is $N_{part}$ for the geometric Glauber model. As described above an exclusivity condition $\Delta z_\text{off}$ is also applied to determine the minimum interval from one \pn\ {\em collision} to the next, from one ``true'' participant to the next according to the the exclusivity model. The sum is $N_{part}$ for that model.

\begin{figure}[h]
  \includegraphics[width=1.68in]{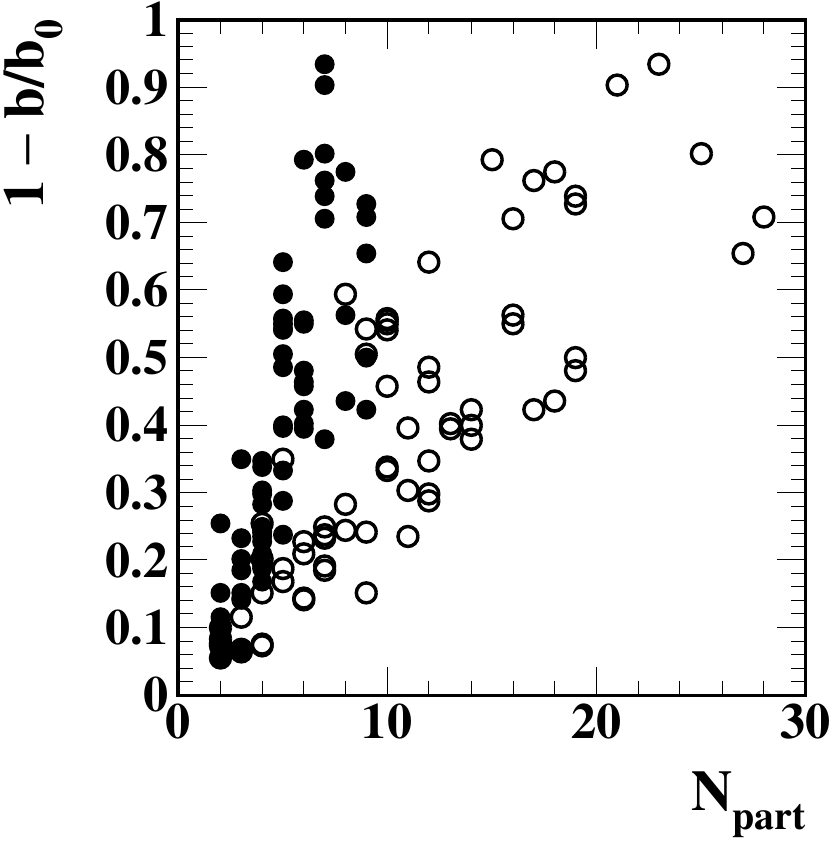}
  \includegraphics[width=1.62in]{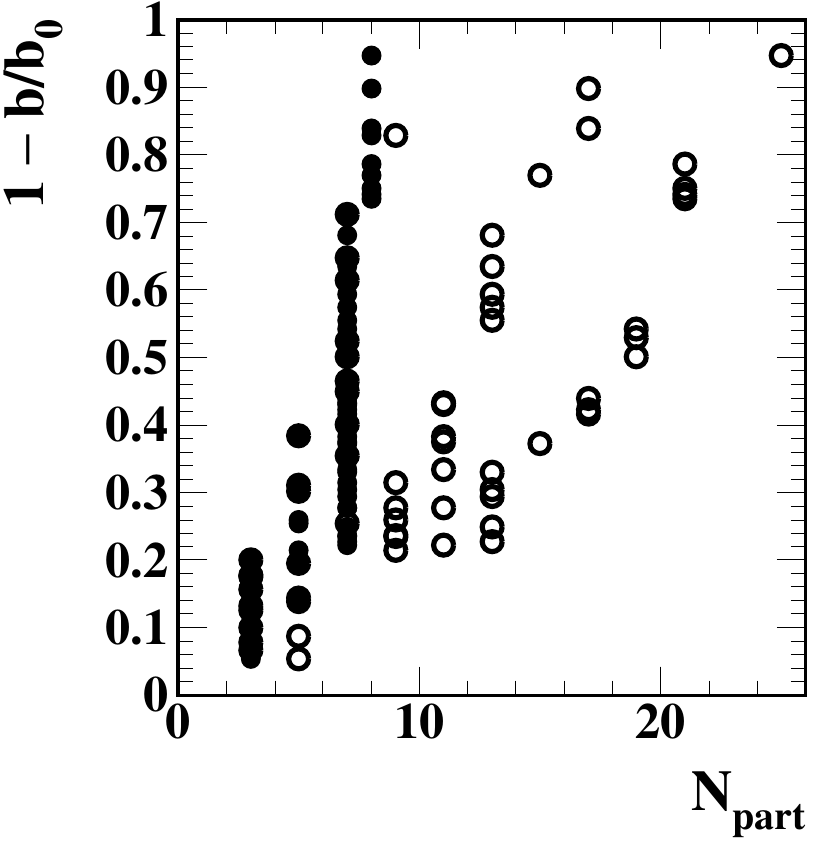}
  \caption{\label{glauber5}
  Left: Participant number $N_{part}$ vs \ppb\ centrality in the form $1 - b/b_0$ for the geometric Glauber MC (open circles, encounters) and the exclusivity model (solid dots, true participants). The simulated Pb nucleus consists of a random nucleon distribution (overlapping nucleons) with mean density 0.14/fm$^3$.
Right: As at left  but for a periodic Pb nucleus with the same mean density and no nucleon overlap.
   } 
\end{figure}

Figure~\ref{glauber5} (left) shows $N_{part}$ vs \ppb\ centrality in the form $1 - b / b_0$ for the geometric Glauber model (open circles) and for the exclusivity model (solid dots). Although the scatter in the data is substantial (due in part to Poisson density fluctuations) the general trends are consistent with Fig.~\ref{glauber22} (right). A high-statistics geometric-Glauber simulation shows encounters extending out to 30 or more~\cite{aliceppbprod}, whereas the maximum number of participants for central \ppb\ collisions is $\approx 10$ for $\Delta z_\text{off} \approx 1.4$ fm (compare to nucleon diameter $\approx$ 1.7 fm or \pp\ impact parameter $b_{0pp} = 1.5$ fm corresponding to $\sigma_{pp} = 70$ mb).

Figure~\ref{glauber5} (right) shows the same procedure applied to a simulated {\em periodic} Pb nucleus with the same mean density but no nucleon overlap. Although nucleon periodicity leads to the band structure the general results are again consistent with Fig.~\ref{glauber22} (right). The large difference does not result from details of Pb nucleus structure.

\section{Summary}

Determination of \ppb\ centrality by two methods leads to very different results. A geometric Glauber model based on the eikonal approximation applied to 5 TeV \ppb\ collisions (and to \pn\ collisions)  assuming that produced hadron density is proportional to number of nucleon participants $N_{part}$ (with \pn\ binary collisions $N_{bin} = N_{part} - 1$) leads to $N_{part}$ estimates greater than 20 and hadron multiplicity per participant pair near 5 for central collisions. Estimates based on a two-component (soft + hard) model (TCM) of hadron production applied to ensemble-mean transverse momentum \mmpt\ lead to $N_{part}$ less than 8 with mean hadron multiplicity per participant pair increasing to about 30 for central \ppb\ collisions.

The TCM applied to \pt\ spectrum and correlation data from \pp\ collisions reveals that the hard component (dijet production) increases as the {\em square} of the soft component (proton dissociation) whereas exponent 4/3 would be expected according to the eikonal approximation. The TCM result can be interpreted to conclude that for \pp\ collisions centrality (impact parameter) is not relevant and that each participant parton in one proton can interact with any parton in the partner proton: a \pp\ collision is ``all or nothing'' for nucleon constituent partons.

Within the TCM context these \pp\ and \pa\ results can be combined to conclude that \pn\ interactions within \pa\ collisions are {\em exclusive}: The projectile proton can interact (collide) with only one target nucleon within a certain distance (comparable to the proton diameter), although other target nucleons may overlap the same volume. \pn\ exclusivity limits the total number of participant target nucleons to 8 or less even for central collisions. If correct that conclusion has significant implications for collision models based on the eikonal approximation and for interpretations of data based on a geometric Glauber model.


\end{document}